# PORTING OF EPICS TO REAL TIME UNIX, AND USAGE PORTED EPICS FOR FEL AUTOMATION

T.V. Salikova

Budker Institute of Nuclear Physics, 630090, Novosibirsk, RUSSIA

## Abstract

This article describes concepts and mechanisms used in porting of EPICS (Experimental Physical and Industrial Control System) codes to platform of operating system UNIX. Without destruction of EPICS architecture, new features of EPICS provides the support for real time operating system LynxOS/x86 and equipment produced by INP (Budker Institute of Nuclear Physics). Application of ported EPICS reduces the cost of software and hardware is used for automation of FEL (Free Electron Laser) complex.

## PORTING OF EPICS TO REAL TIME UNIX

The aim of the porting [1] was the reduction of cost of hardware and software required for creation of distributed control system which is built on base of personal computers with processors Intel x86, which has best relation of the cost per performance. INP has the rich experience of development of the special equipment for the particle accelerators, which is used for creation of FEL [2], and CAMAC devices made in INP is applied for automation of FEL. LynxOS/x86 [3] was chosen as low-cost UNIX real time operating system that supports POSIX standards. Besides, free software of GNU and XFree86 projects works at LynxOS platform.

Release 3.13.2 was chosen as robust variant of EPICS [4] for FEL control system, but the porting was started with release 3.13.0.beta12 [5] in 1998. Now I do not use R3.14, which supports Operating System Independent (OSI) features, by two reasons: first, I have not time for the preparation OSI package for LynxOS and testing of it. I am afraid that R3.14 requires several corrections of the device support routines and CAMAC driver. Second, FEL control system would be fully installed by the end of this year.

The porting of EPICS has two stages:
a) support of LynxOS/x86 at OPerator Interface (OPI),
b) libraries emulates VxWorks [6] calls which is used by Input Output Controller (IOC). Writing of CAMAC driver that supports synchronous and asynchronous requests.

The inclusion of LynxOS/x86 support at OPI level is trivial procedure. It required the preparation of configuration files. Channel Access (CA) package of these EPICS versions is built on 4.3BSD socket system, then LynxOS/x86 2.5 and later versions uses 4.4BSD. Only R3.13.3 give opportunity of CA compilation as 4.3BSD and 4.4BSD. It requires little corrections in the routines of CA library working with *ifreq* structure [7].

The general requirement of support LynxOS at IOC level is conservation of EPICS source without modification. For this purpose, the list of VxWorks calls was prepared, list contains all calls used by IOC. On base this list, libraries were created that was fully emulated VxWorks calls. Multitask mechanism of VxWorks is reconstructed by multithreads features of LynxOS easy.

1) Ring buffer subroutines library *rngLib.h* is emulated by the set of routines which manages FIFO circular buffer, which is described *RING* structure such as VxWorks uses. Routines of linked list library *lstLib.h* was fully copied from VxWorks.
2) General semaphore library *semLib.h*, binary semaphore library *semBLib.h*, counting semaphore library *semCLib.h*, mutual-exclusion semaphore library *semMLib.h*. For emulation of different types of semaphores was chosen counting semaphores of LynxOS. All semaphore identifiers is described by *SEMAPHORE* structures that is placed in doubly-linked list. This list is processed by emulated routines as *semDelete, semTake, semGive, semFLush* and other. *SEMAPHORE* structure contains information about type of semaphore, priority, counter, timeout and additional information about threads that works with this semaphore.
3) Task information library *taskInfo.h*, task management library *taskLib.h*. The set of these calls is emulated by POSIX draft 4.9 multithreads features of LynxOS. LynxOS thread is equivalence VxWorks task. All task identifiers *tid* is placed into ring buffer, that is dispatched by special *pthread*. This *pthread* controls state of all spawned tasks.

4) Watchdog timer library *wdLib.h*. Watchdog timer mechanism is emulated by the creation of POSIX interval timer for the calling process [8]. The extended *WDOG_ID* structure describes POSIX timer and his queue.

The problem of symbol table subroutines library is not solved, *iocCore* is compiled with libraries that emulate VxWork features, libraries of record support routines and of CAMAC device support routines fully. The necessary entry points of record support routines and device support routines is enumerated in *src/db/iocInit.c* file. Sequencer does not works of this reason. The IOC structure is shown on Fig. 1.

*iocCore* has the possibility of launching of X11 subwindow with several IOC test routines: *dbl, dba, dbpr, scanppl, scanpel, scanpiol* (the list of routines will be extended).

On the analogy of VME/VXI, CAMAC, BITBUS and other, the support of CAMAC_PPI bus is added to *src/dbStatic/link.h* and *src/dbStatic/dbStaticLib.c* files. CAMAC_PPI interface is ISA card with six channels for the connection of crate controllers. This IOC supplement serves CAMAC records, which has next capabilities: the maximal number of NAF for record initiation is 5, the maximum of NAF for record processing is 9. The chain of NAF consists of following information: N – the crate number, A – subaddress, F – function, data, cycle counter for waiting Q signal and time interval in ticks for delay in this cycle.

Due to CAMAC_PPI supplement in *src/dbStatic/dbStaticLib.c*, Database Configuration Tool *dct313* supports the preparation of records working with CAMAC devices. *dct313* allows preparation of a block of NAF. CAMAC devices is serviced by the set upward of 20 device support routines that were written for optimization of exchange with equipment.

## USAGE PORTED EPICS FOR FEL AUTOMATION

The distributed control system of FEL is built on private subnet providing minimal traffic. All control computers is connected to "Ethernet segment" which attached to the gateway. There are personal computer under supervision of Linux with IP-masquerade. All control computers has the access to Internet, but there is not external access to them.

The high performance Pentium allows one to join OPI and IOC functions in one computer, the average performance is equal to 4000 records per second at Pentium-III 550MHz, which simultaneously services OPI and IOC features of Radio Frequency (RF) control system. Test executes *get* and *put* requests to *ai* and *ao* records in *iocCore,* servicing RF database consisting of 483 records. Those are usual debugging conditions for the adjustment of FEL subsystem.

At present time, FEL complex can be divided into four autonomous subsystem, each having its own IOC:

a) RF system joins three cavities and three generators of injector and sixteen cavities and two generators of microtron. RF IOC services 481 records.
b) Injector IOC controls electron gun and solenoids, lenses of injector and beam transducers. Database contains upward of 150 records.
c) IOC of microtron magnetic system controls about 300 parameters of power supplies of bend magnets, solenoids, quadrupoles and undulators.
d) Diagnostic IOC services diagnostic equipment and checks temperatures in 120 points of microtron.

The adjustment of RF IOC and Injector's IOC is being finished. Now last two IOC is designed.

OPI programs is compiled with libraries of Graphical User Interface (GUI) which is built on free software: XFree86 4.1.0, OpenMotif 2.1.30, TCL/TK 8.4 tool kits, which supports modern video cards and monitors as well as new features of graphical libraries. The software choice was defined by possibility of upgrade of video resources in future.

OPI programs of FEL control system is built by two means:

a) control panels for operator-physicist is set of subwindows that is spawned by widgets as *menu_pull* at menu bar or *PushButton*. Subwindows contains tables and graphs of input data. Set of *scale*, *SpinBox* and other widgets is used for output data. FR console is created in this way.
b) control panel of program for government of injector (also another subsystem of FEL) has mnemonic diagram of elements of FEL injector. It is similar to automations symbols on panels of industrial automation applications made by National Instruments [9]. Each element on panel is linked with its subwindow. Clicking on a mnemonic element will spawn of subwindow, control panel of this element. Also color palette is used to indicate the alarm status. If device fails, then corresponding element in the mnemonic diagram is painted in red.

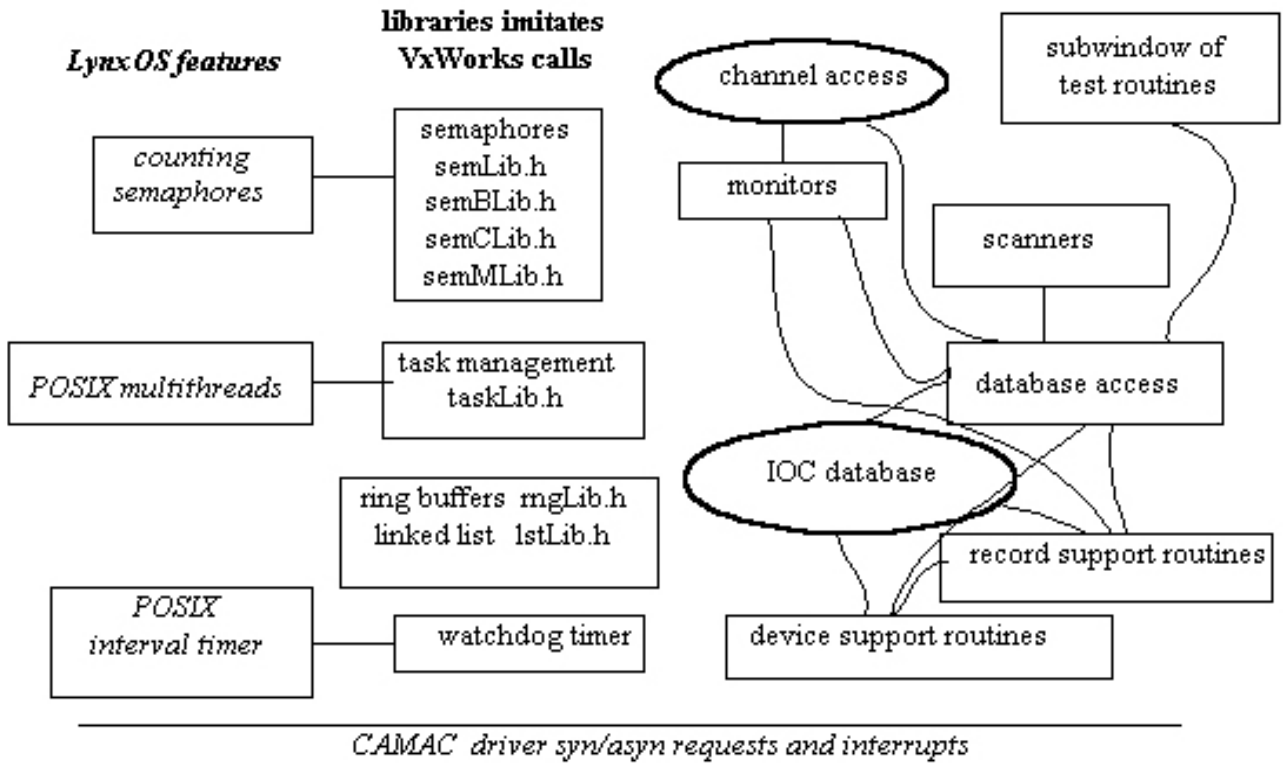

Fig.1 The structure of *iocCore*.

## CONCLUSIONS

The EPICS tool kit provides environment for development of the FEL control system in the modern state of computer technologies. The ported version of EPICS allows utilization of hardware made by INP, which reduces cost of hardware used for FEL automation. The average performance of IOC at Pentium III (550 MHz) is equal to 4000 records per second, which is enough to make a good control system.